# Alternative difference analysis scheme combining R-space EXAFS fit with global optimization XANES fit for X-ray transient absorption spectroscopy

Authors

**Fei Zhan[ab], Ye Tao[a] and Haifeng Zhao[a]\***

[a]Beijing Synchrotron Radiation Facility, Institute of High Energy Physics, Chinese Academy of Sciences, Beijing 100049, People's Republic of China

[b]University of Chinese Academy of Sciences, Beijing, 100049, People's Republic of China

Correspondence email: zhaohf@ihep.ac.cn

**Funding information**          National Natural Science Foundation of China (award No. U1332205); Knowledge Innovation program of the Chinese Academy of Sciences (award No. KJCX2-W-N42).

**Synopsis**  We developed a difference analysis scheme for the time-resolved X-ray absorption spectroscopy. It combines the R-space difference EXAFS fit with the difference XANES fit, characterized by changeable calculation core and global optimization algorithm. It was applied on a photo-excited spin crossover iron complex.

**Abstract**          Time-resolved X-ray absorption spectroscopy (TR-XAS), based on laser-pump/X-ray probe method, is powerful in capturing the change of geometrical and electronic structure of the absorbing atom upon excitation. TR-XAS data analysis is generally performed on the laser-on minus laser-off difference spectrum. Here we present a new analysis scheme for the TR-XAS difference fitting in both the Extended X-Ray Absorption Fine Structure (EXAFS) and the X-ray Absorption Near Edge Structure (XANES) region. R-space EXAFS difference fitting could quickly give main quantitative structure change of the first shell and provide reliable constraint on the range of the variables in XANES fit. XANES fitting part introduces global non-derivative optimization algorithm and optimizes the local structure change in a flexible way where both the core XAS calculation package and the search method in the fitting shell are changeable. The scheme was applied to the TR-XAS difference analysis of Fe(phen)$_3$ spin crossover complex and yielded the reliable distance change and the excitation population.

**Keywords: difference XAFS fit, global optimization, R-space, core/shell substitution, TR-XAS**

## 1. Introduction

Time-resolved X-ray absorption spectroscopy (TR-XAS), based on laser pump/X-ray probe method, is a powerful probe to address geometric and electronic structure of light generated transient species (Chen *et al.*, 2014; Chergui, 2016). Its time resolution capability is dependent on X-ray pulse duration. Generally, it can reach 100ps temporal resolution using synchrotron radiation sources and less than 100 fs using X-ray free electron lasers. The TR-XAS has been widely applied in photochemistry and photophysics fields. Bressler studied light-induced spin crossover of an iron complex (Bressler *et al.*, 2009). Zhang obtained 0.01 Å high resolution structure change of an excited state of osmium complex (Zhang *et al.*, 2014). Moonshiram revealed intermediate molecule structure of Co(II) and Co(I) photocatalyst in a real $H_2$ production (Moonshiram *et al.*, 2016). Wen studied $BiFeO_3$ nanofilm and found anisotropic unit cell change of in-plane contraction and out-plane extension (Wen *et al.*, 2015).

Benefit from the well developed X-ray Absorption Fine Structure (XAFS) spectrum analysis methods in the last forty years, kinds of TR-XAS treatments or packages have been developed both in EXAFS and XANES region along with TR-XAS extensive applications. The differential EXAFS fitting method in q-space is applied by Borfecchia to study the photoactive metal complexes cis - [Ru(bpy)$_2$(py)$_2$] (Borfecchia *et al.*, 2014; Garino *et al.*, 2014). Zhang et.al adopted a fit of excited state EXAFS reconstructed according to different excitation fraction (Zhang *et al.*, 2015). FEFF, a popular XANES computation package based on the self-consistent multiple scattering theory, is used in the study of PtPOP system by Van der veen, who took Kas' Bayes XANES fit method to search for the structure change information when the system is excited (van der Veen *et al.*, 2010). Interpolation approach is effective in structure refinement, Smolentsev et al took it to invoke FEFF and FDMNES in the TR-XAS fitting research on platinum dimer (Lockard *et al.*, 2010). They also used DFT-MO method to simulate pre-edge feature of the different cobalt species (Moonshiram *et al.*, 2016). MXAN (Benfatto *et al.*, 2003), one of the widely used XANES fit algorithms, takes CONTINUUM to do the XAFS calculation and adopts MINUIT optimization package. MXAN has been applied on the characterization of molecule excitation state of [Ru$^{II}$(bpy)$_3$]$^{2+}$ by Benfatto (Benfatto *et al.*, 2006).

Here we present an alternative data analysis scheme for the TR-XAS. It combines R-space EXAFS fit with global optimization XANES fit for the TR-XAS analysis. The fit of the difference spectrum is done directly in both EXAFS and XANES region. We adopt R-space fit to avoid one more reverse Fourier Transform (FT). The XANES difference spectrum fit is similar to that of MXAN, but it is more flexible, as we can choose different calculation cores, like FEFF (Rehr *et al.*, 2010) or FDMNES (Joly *et al.*, 2009) and different adaptive optimization algorithms as well to control the fit process.

## 2. Method

### 2.1. Workflow

We present here a TR-XAS fit scheme in both EXAFS and XANES region where the R-space difference fit in EXAFS gives the average bond length changes of first shell around the absorber, while the detailed structural change is derived in XANES difference fit. Our scheme is flexible in XANES fit where both the search algorithm (shell) for variables and the spectrum calculation package (core) could be substituted by any suitable packages.

The workflow of our TR-XAS fit scheme composes of two blocks, as shown in Fig.1 the EXAFS difference fit (block 1) and Fig. 2 the XANES difference fit (block 2), each one starts from the finding of ground state structure and is followed by the seek for the structural change of intermediate state. The spectrum of ground state or intermediate state is first calculated with general EXAFS formula (Newville, 2013) in block 1 or with selected XAS calculation package in block 2. The following theoretical difference spectrum is obtained according to the equation,

$$\chi_{th\_diff} = \alpha(\chi_{IntS} - \chi_{GS}) \quad (1)$$

where $\alpha$ is the fraction of the intermediate state, $\chi_{GS}$ and $\chi_{IntS}$ are the theoretical spectra of ground state and intermediate state, respectively. The theoretical difference spectrum $\chi_{th\_diff}$ is compared with the experimental difference spectrum $\chi_{exp\_diff}$ to check their consistency in block 2. They are, however, firstly Fourier transformed into R-space in block 1 before comparison. The structure of intermediate state and $\alpha$ may be modified by parameter search shell to do the difference spectrum calculation again, following the comparison with experimental spectrum until the consistency is acceptable.

We give here the definition of goodness of fit, the minimum of which in the space of parameter is what the difference fit searches for:

$$R(x_j | j = 1, n) = n \frac{\sum_{i=1}^{m} w_i \left\{ \alpha * \left[ (\chi_{IntS}^i - \chi_{GS\_th}^i) - \chi_{exp\_diff}^i \right] \varepsilon_i^{-1} \right\}^2}{\sum_{i=1}^{m} w_i} \quad (2)$$

where $x_i$ is the parameter to be fit, n is the number of independent parameters, m is the number of data points, $\varepsilon_i$ is the individual error in the experimental data set, and $w_i$ is the statistical weight.

When $w_i$ is one, the function reduces to the statistical $\chi^2$ function.

The conclusion of difference fit in EXAFS region is helpful for the fit in XANES, as one could limit the varying range of some parameters to save the fit time.

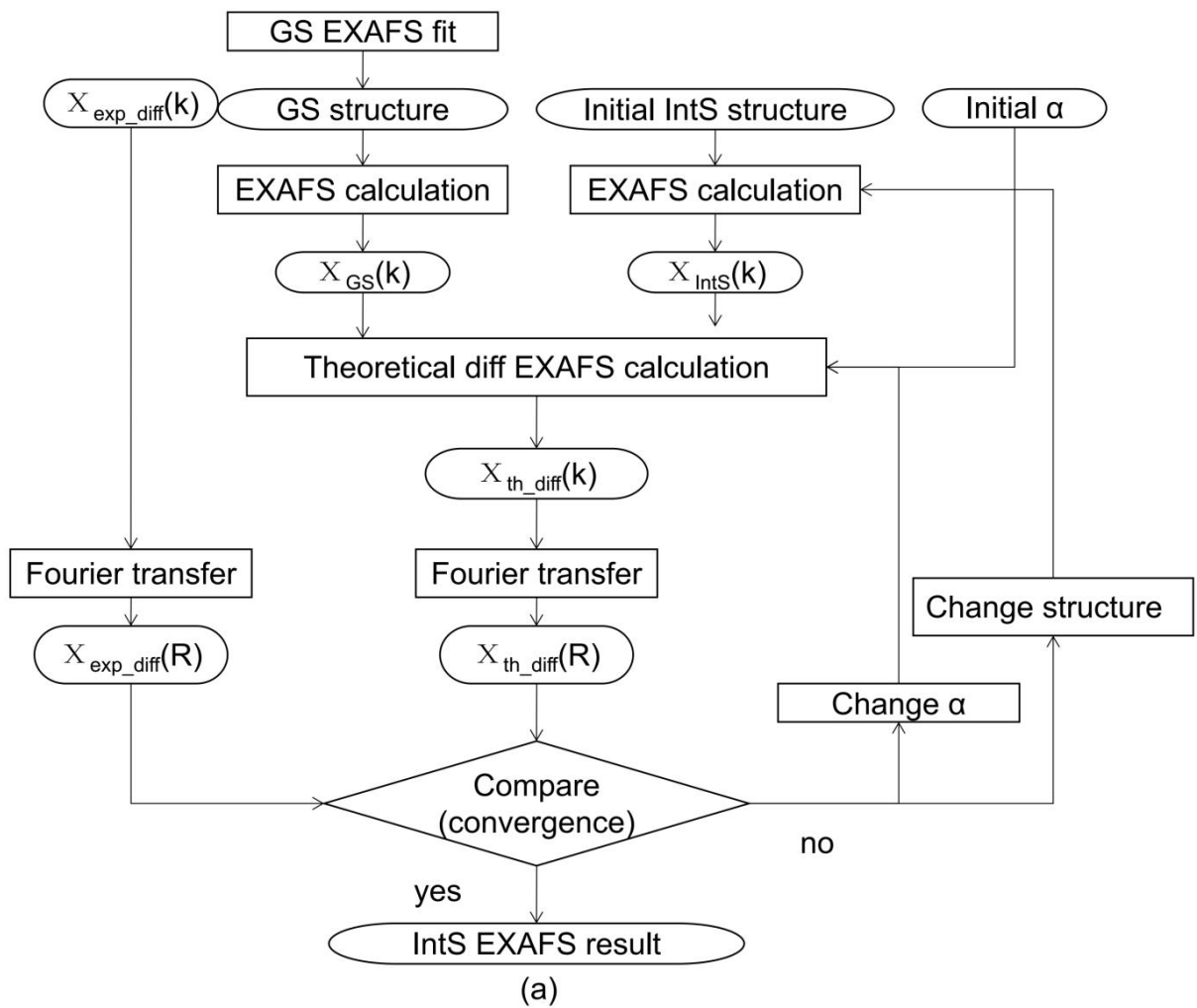

(a)

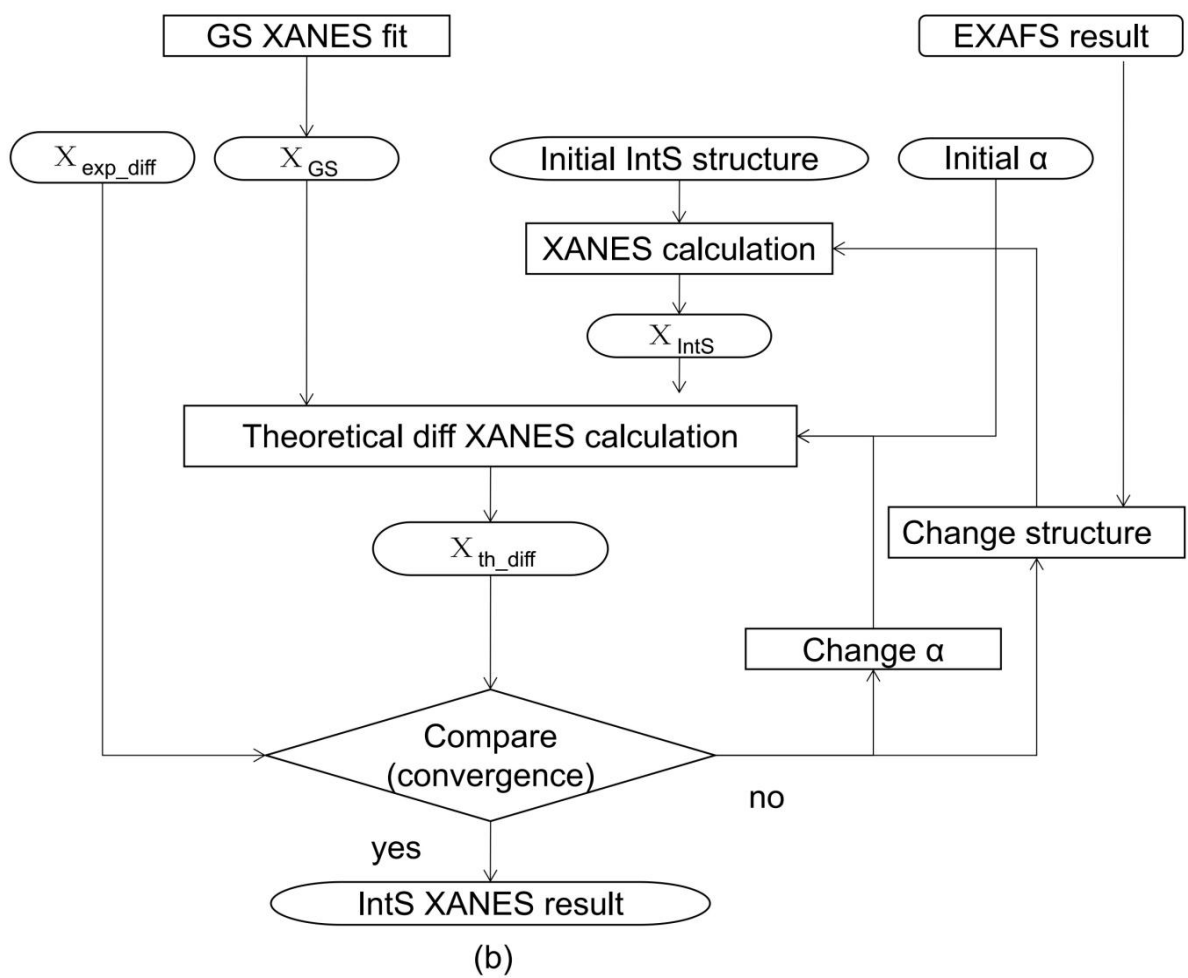

(b)

**Figure 1** (a): The flowchart of EXAFS difference fit, 'GS' stands for ground state, 'IntS' for intermediate state, 'diff' for difference, 'th' for theoretical and 'exp' for experimental. The data operation is presented in square frame while the input/output of operation are donated in the elliptical frame. (b): The flowchart of XANES difference fit. The donations and interpretation of frame are same as those in (a).

TR-XAS data analysis is generally performed on the laser-on minus laser-off difference spectrum. The difference highlights the structural change, since the difference spectrum eliminates the experimental system error. Our framework does the fit directly to the difference spectrum instead to the ground state and the excitation state spectrum separately. Moreover, the pre-process of spectrum in EXAFS such as the background subtraction is no need in difference fit as they are same in ground state and intermediate state. Obviously we can see in equation (1) the fraction of intermediate state $\alpha$ in the system after pump is critical in the produce of XANES fit. There are several ways to estimate $\alpha$ in TR-XAS analysis. Moonshiram *et al.* get the estimation of Co(II) intermediate fraction via using Co(II) reference sample's pre-edge fingerprint (Moonshiram *et al.*, 2016). It would fail when the analogue sample is hard to synthesize or is unstable in the experiment, which occurs in most cases. Some groups obtain the fraction from optical transient absorption (OTA) experiment measured under the same condition as the TR-XAS (Benfatto *et al.*, 2006; Borfecchia *et al.*, 2013), but some system is OTA silent. One general choice to get the intermediate fraction is treating it as a variable like other structural parameters in the TR-XAS fit (Lockard *et al.*, 2010; Zhang *et al.*, 2015), as we do in our scheme.

We take R-space instead of k-space to do the EXAFS fit. It is hard to get the high quality TR-XAS signal up to full EXAFS range. Moreover, it is clear and straightforward to find the trivial difference between simulation $\chi(R)$ and experimental one in R-space. Generally the major structure change in the system comes from the nearest atoms around the absorber, say the first shell, we can limit the fitting domain to the first shell in R-space during the fit, the calculation of $\chi$ will be much fast as we don't need to consider the scattering from other shells.

In XANES fit, as described above, the core simulates the spectrum and the shell controls the parameter search. FEFF (Rehr *et al.*, 2010) is taken to do XANES calculation by default in the scheme, and it could be replaced by FDMNES (Joly *et al.*, 2009) at present. The search method adopted by shell is also substitutable, such as the Mesh Adaptive Direct Search (MADS), the dividing rectangles method (Gablonsky & Kelley, 2001) and improved stochastic ranking evolution strategy (Runarsson & Yao, 2000). We take here MADS method as default for the difference fit. It evolved from of the Generalized Pattern Search which is one of the modern grid searches. MADS has been proved to be efficient (Audet *et al.*, 2008, 2011),and is now widely used (Miiller *et al.*, 2012; Eiswenhoer *et al.*, 2012; Berrocal-Plaza *et al.*, 2014). Moreover, NOMAD, which is one of the package using MADS method, offer the parallel ability (Audet *et al.*, 2008).

## 2.2. Choice of the optimization algorithm

Optimization algorithm is important in the XANES fit. Traditional structure optimization in computation chemistry adopts and develops series of local optimization algorithms with initial hessian guess method and hessian update method. However, in crystal structure prediction area, the application of heuristic global optimization algorithms, the particle swarm optimization in CALYPSO code (Wang *et al.*, 2012)and the evolutionary optimization in USPEX code (Glass *et al.*, 2006), has archived great success. The choice of optimization algorithm also should be deliberated in XANES fit. First we prefer to get global minimum rather than local minimum. Second, since XANES fit is a black-box optimization, we don't know whether it is convex problem before the fit, so generally we adopt nonlinear optimization algorithm. Moreover, we decide to use non-derivative algorithms, as XANES fit also can't provide analytical derivative, and the cost of numerical derivative computation of XANES fit is huge, also the reliability of numerical derivative is hard to be guaranteed. Fortunately many global optimization algorithms don't need information of derivative, so global non-derivative algorithm is our choice. Deterministic algorithm, stochastic and heuristics algorithm are two main branches of global optimization. Normally the deterministic algorithm has theoretical convergence analysis under the given condition, while some stochastic and heuristics algorithms can only provide "probabilistic convergence guarantee" or " remain heuristics " (Pardalos *et al.*, 2000). In deterministic algorithm, taking mesh adaptive direction search (MADS) as an example, Audet *et al* gave MADS's hierarchy of convergence analysis (Audet & Dennis Jr, 2006), and proved that the lower triangular instance of MADS can produce dense set of poll directions. Another example is DIRECT algorithm. It can prove its ability of convergence to Karush-Kuhn-Tucker point under mild condition (Gablonsky & Kelley, 2001). Though some optimization algorithms have convergence analysis under given condition, the theoretical condition is usually hard to test. Therefore the benchmark of the algorithm is important in practice. From the NLOPT library's benchmark (Kumar *et al.*, 2016) and the derivative-free algorithms' benchmark (Rios & Sahinidis, 2013), we can find that usually one optimization algorithm can't solve all test problems. Also there is an optimization solution distribution of multiple optimization solver runs due to algorithm itself or initial solution. Obviously, there is no single ultimate algorithm for all optimization problems, so the substitutable algorithm in shell of XANES fit is a good choice.

## 3. Application

### 3.1. Sample and experimental description

We applied our scheme on a model spin crossover iron complex, 1,10-Phenanthroline iron(II) sulphate, Fe(II)(phen)$_3$. Fe(II)(phen)$_3$ is a low-spin (LS) complex in ground state, and it becomes high spin (HS) state upon photo-excitation due to the spin crossover. It is important in dynamic magnetic research and has been well studied by the TR-XAS . Nozawa group reconstructed the EXAFS of the excited state of Fe(II)(phen)$_3$ and found 0.17Å bond elongation for the first shell comparing to the low spin ground state (Nozawa *et al.*, 2010). Fe(II)(phen)$_3$ was purchased from Alfa Aesar and was used without any further purification. The HS analogue complex, Fe(II)(2-CH$_3$-phen)$_3$ was synthesized. Its HS state results from steric hindrance of the methyl. Fe(II)(2-CH$_3$-phen)$_3$ not only provides a HS analogue for structure analysis but also helps to determine the excitation fraction.

The TR-XAS measurement was performed at the beamline 11-ID-D of the Advanced Photon Source. The experimental description could be found in details elsewhere (Chen & Zhang, 2013). The Fe(II)(phen)$_3$ laser-on and laser-off spectra are shown in Fig. 2 along with XAS of Fe(II)(2-CH$_3$-phen)$_3$.

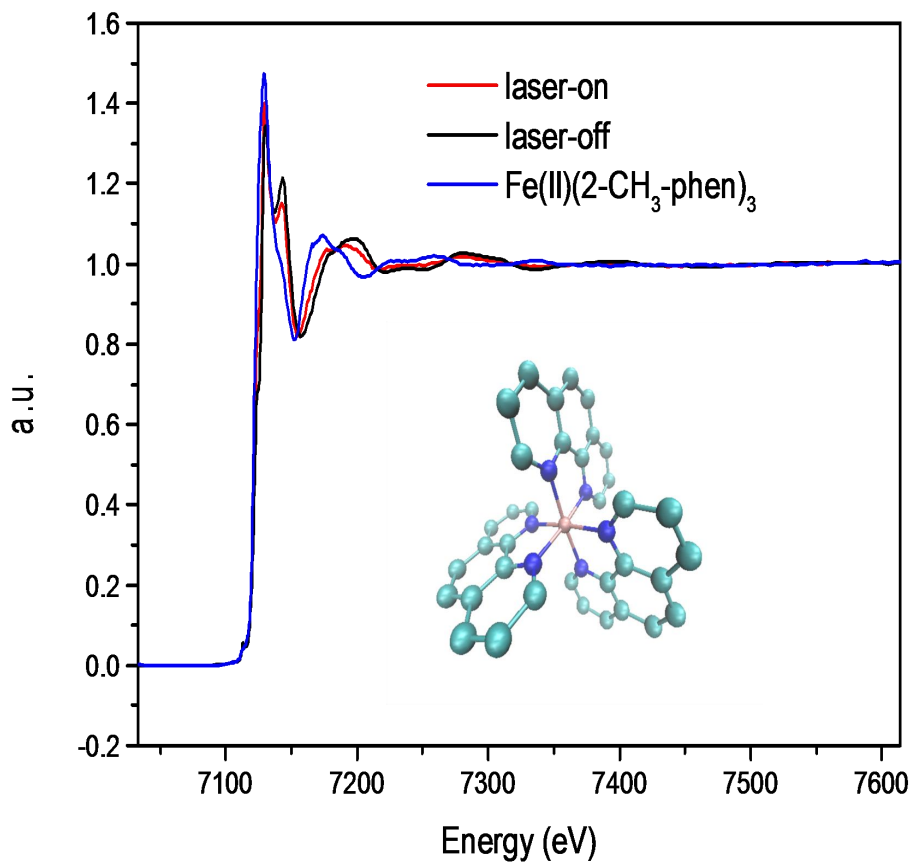

**Figure 2** The TR-XAS of Fe K-edge of the Fe(II)(phen)₃, the laser-on (red) and laser-off(black) along with XAS of F*e*(II)(2-CH₃-phen)₃(blue). The TR-XAS was measured in fluorescence mode. The laser excitation wavelength is 400 nm with repetition rate 10 kHz. Inset: Molecular structure of the Fe(II)(phen)₃, the iron atom is coordinated to six nitrogen atoms in octahedral configuration, and each phenanthroline ligand provides two nitrogen atoms for coordination.

In EXAFS fit, We fit the ground state Fe(II)(phen)₃ and its HS analogue Fe(II) (2 -CH₃ -phen)₃. The data was processed by our scheme with EXAFS equation referring to the one in Larch (Newville, 2013). The amplitude reduction factor is fixed to 0.9, and the coordination number is fixed to 6. The variables are the disorder factor $\sigma^2$, the energy shift $E_0$ and the bond length r. The theoretical amplitude and phase are extracted from the 1.95 Å Fe-N single scattering path calculated by FEFF9.

The R-space fit is displayed in Fig. 3 and the fit results are listed in Table 1. The disorder factor and energy shift will be used in the following difference fit.

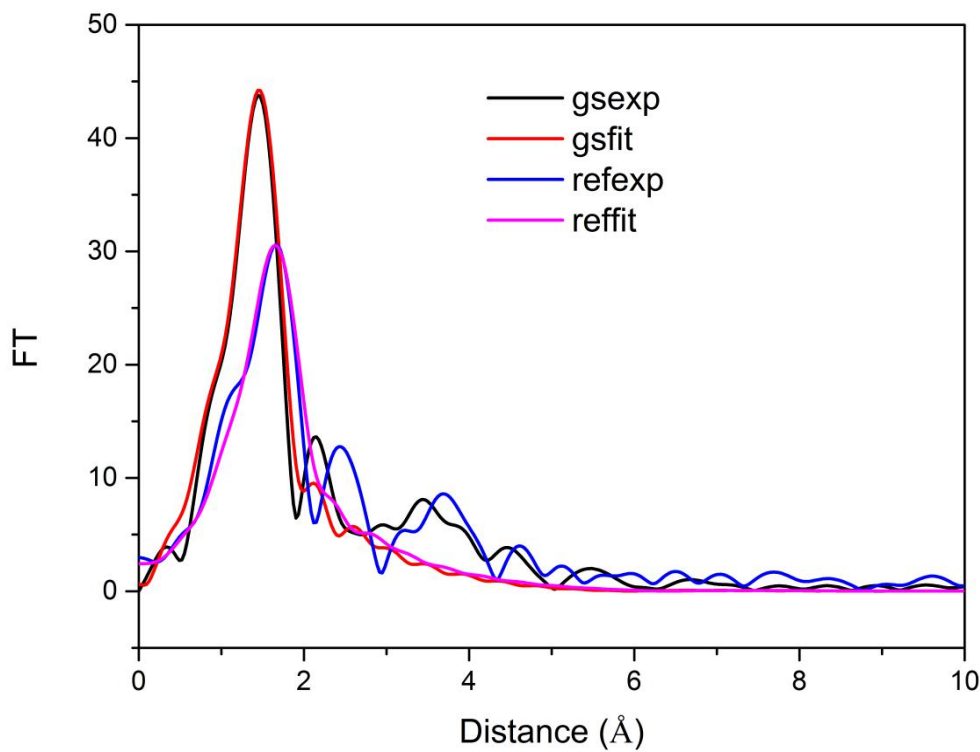

**Figure 3**  Experimental and simulated Fourier Transforms of $k^2$-weighted EXAFS of ground state(gsexp,gsfit) and reference sample Fe(II) (2 -CH$_3$ -phen)$_3$ (refexp,reffit).

**Table 1**  Fit results of ground state and HS analogue. $r$ is the bond length change related to the reference Fe-N scattering path. $\sigma^2$ is the disorder factor, $E_0$ is the energy shift.

| Sample | $r$ | $\sigma^2$ | $E_0$(eV) | R factor |
|---|---|---|---|---|
| Fe(phen)$_3$ | 1.97 | 0.009 | -4.99 | 0.005 |
| Fe(II)(2 -CH$_3$ -phen)$_3$ | 2.17 | 0.014 | 1.81 | 0.004 |

As stated above, the difference between laser-on and laser-off highlights the excite signal. The difference spectrum (black) is displayed in Fig.4, compared with the difference obtained from HS analogue minus ground state. The discrepancy reflects the excitation fraction. The excitation fraction is 37%. The reconstructed difference spectrum and the difference spectrum obtained from the excitation analogue are in good agreement. In the following we will use our new scheme to fit the difference spectrum in both XANES and EXAFS region to obtain the structure change information and excitation fraction.

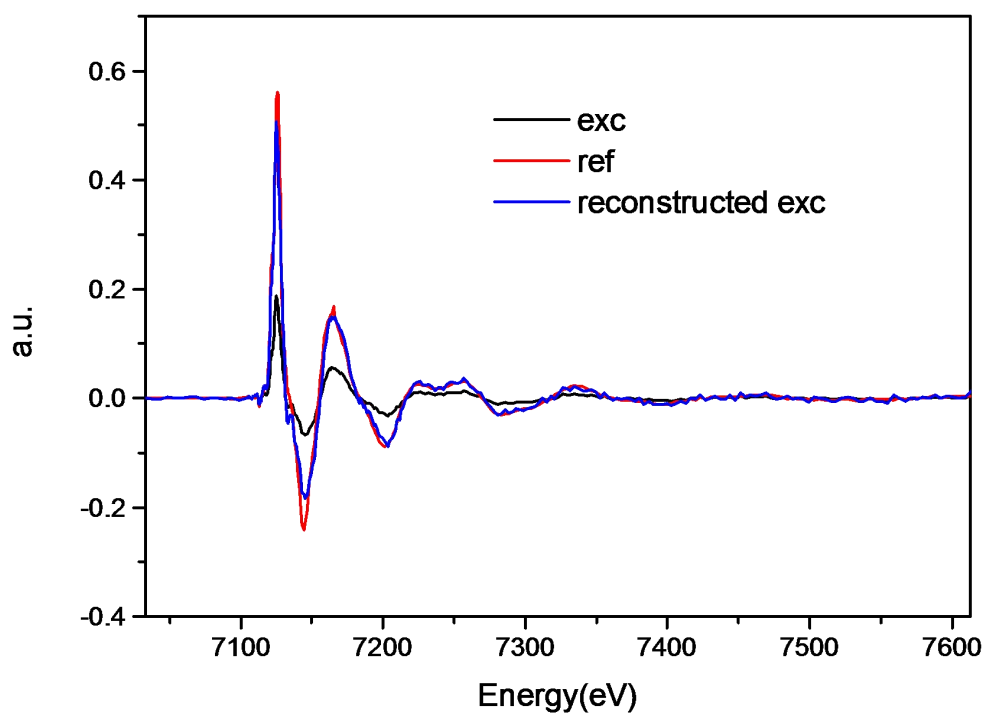

**Figure 4** Difference spectra of the HS excitation state (exc) and excitation analogue reference (ref). The reconstructed difference spectrum (reconstructed exc) is obtained from dividing the experimental one by the excitation fraction 37%.

### 3.2. EXAFS Difference spectrum fitting

The reverse Fourier Transform of the experimental difference spectrum is shown in Fig. 5. We find that 1-1.9 Å R-space window can reveal the main feature of the experimental one. The k-weight of the fit is 2. We use k-space 3-10 Å$^{-1}$, R-space 1-1.9 Å for the fit.

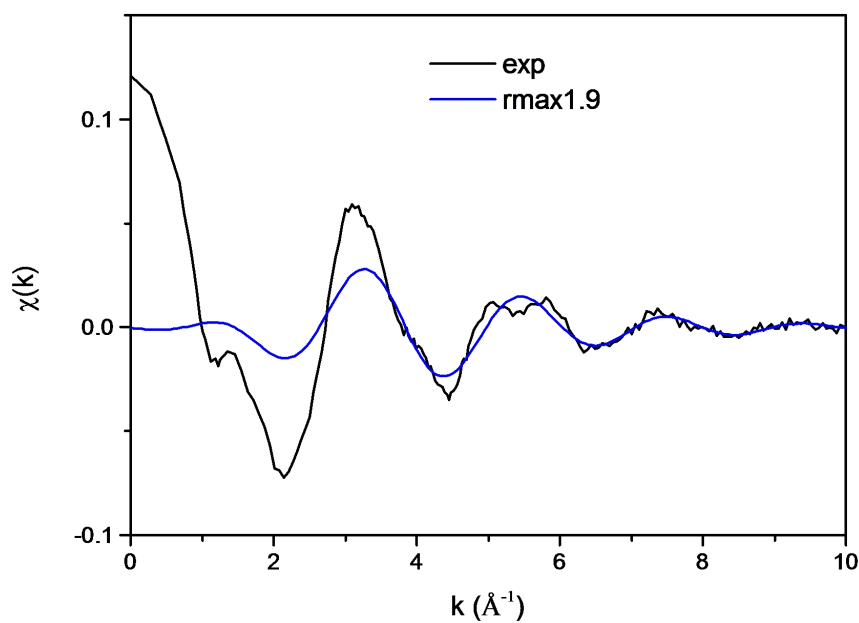

**Figure 5** Experimental and Reverse Fourier Transform of difference spectrum, R-space range is 1-1.9 Å.

We fit the intermediate's fraction and relative bond length change. The energy shift and the disorder factor, given in Table 1, are fixed during the fitting. The R-space fit is shown in Fig. 6(a). We also do the reverse Fourier Transform of the R-space fit result, as seen in Fig. 6(b).The Fe-N bond length is elongated with 0.17 Å under the excitation state. The excitation fraction is 37%. The results are in agreement with those from the HS analogue in Table 2. Difference EXAFS fit can only provide bond length information, so next we perform the difference XANES fitting.

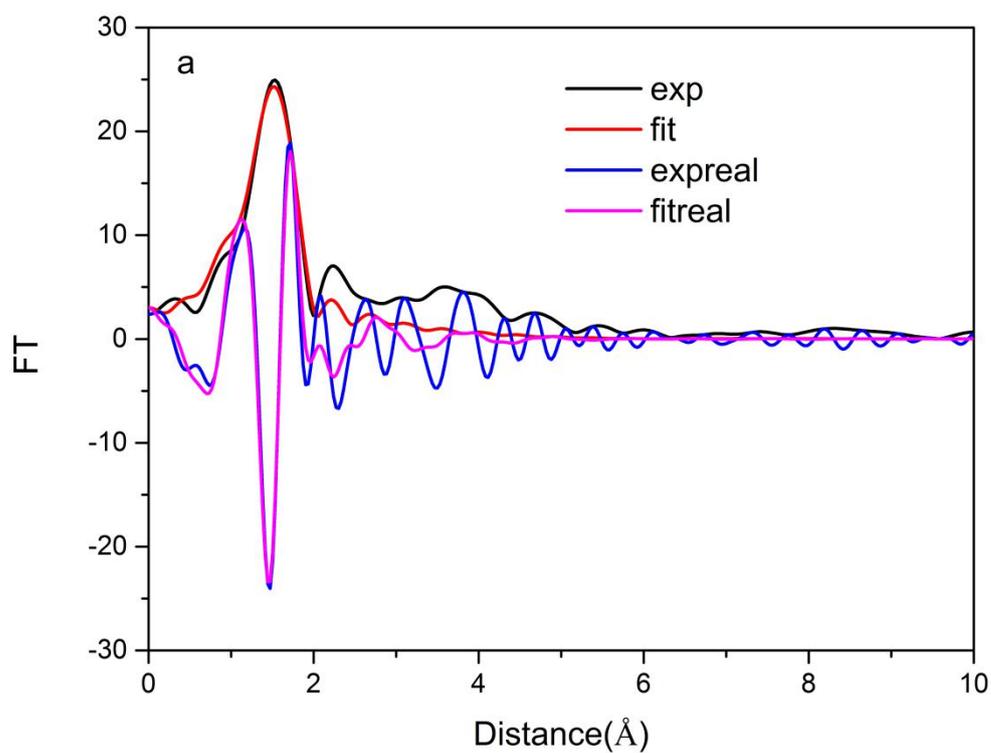

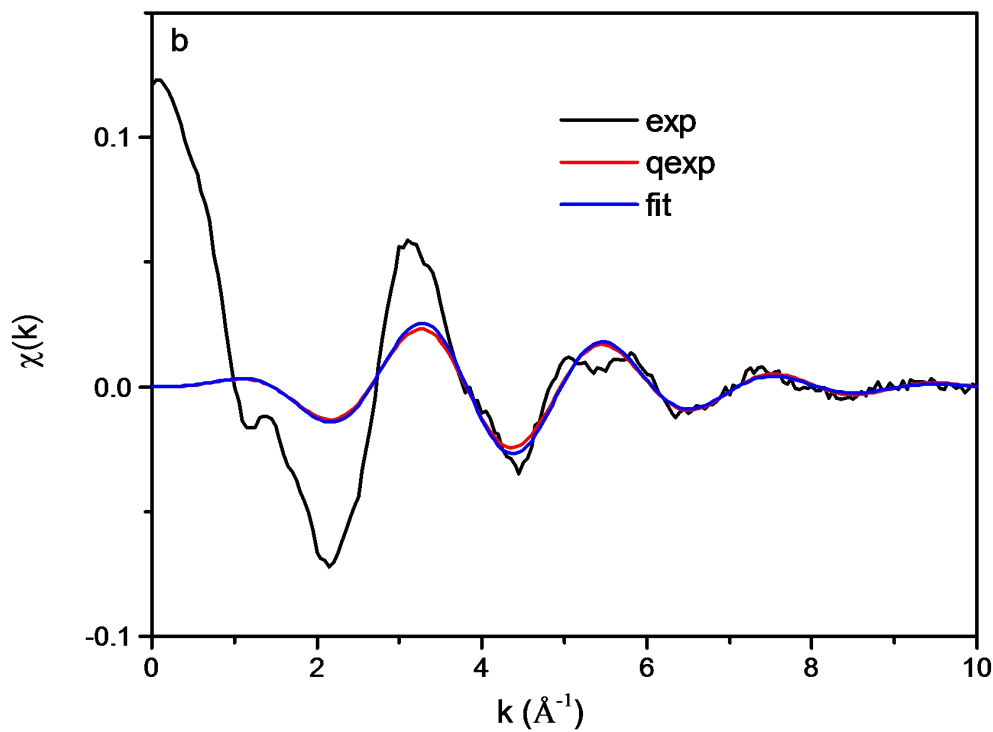

**Figure 6** (a) Experimental(exp) and simulated(fit) Fourier Transforms of $k^2$-weighted difference spectrum. (b) Reverse Fourier Transform of the R-space fit result of the difference spectrum, with the k-space experimental difference spectrum.

### 3.3. XANES difference fit

First we fit the ground state structure. We use FEFF9/FDMNES packages to do the XANES calculation and take NOMAD package to search for the optimized variables. The initial structure of low spin ground state comes from (Yan *et al.*, 2000). The radius of cluster for self-consistent multiple scattering is 6.5Å, containing all of the atoms in the molecule. Real Hedin- Lundqvist exchange-correlation potential is used in the calculation. The phenanthroline ring is taken as a rigid unit during the fit, and we choose the midpoint of the two ligand N atoms to be the representative of this unit. The best fit of FEFF and FDMNES are shown in Fig. 7(a). All the features are reproduced by FEFF and they are much better than those reproduced by FDMNES, so we use FEFF's data here. The average

Fe-N bond length from FEFF is 1.97 Å.

In difference XANES fit, we use FEFF9/FDMNES packages to do the XANES calculation and take

MADS/ISRES/DIRECT_L optimization algorithms to search for the best optimized variables. We fix

the normalization factor, the energy shift of intermediate state to be same as those derived from the

previous ground state fit. The best fit is listed in Table.2. Though FDMNES/NOMAD produces the

lowest R factor, but the spectrum of ground state derived by FDMNES is not good enough. Moreover

its intermediate state fraction is much larger than that from the intermediate analogue and that from

difference EXAFS fit. Checking the R factor of fit given by FEFF combined with different

optimization algorithms in Table 2, the best fit is given by FEFF/NOMAD as shown in Fig.7(b). The

main features in the difference spectrum are reproduced. Coordinates of the intermediate structures

are listed in support information. The average bond length of first shell extents 0.11Å, intermediate

fraction is 39%, as shown in Table 3. The fitting results are close to those obtained in R-space EXAFS

fit and the data reported (Nozawa *et al.*, 2010).

**Table 2**    HS state fit result with different calculation core and search method

| Core | Search method | Bond length change | Fraction | R factor | Structure | Picture of fit result |
|---|---|---|---|---|---|---|
| FEFF | NOMAD | +0.11 | 39% | 0.0058 | Table S1 | Fig.7(b) |
| FEFF | ISRES | +0.12 | 26.1% | 0.0096 | Table S2 | Fig.S1 |
| FEFF | DIRECT_L | +0.08 | 44.2% | 0.0066 | Table S3 | Fig.S2 |
| FDMNES | NOMAD | +0.10 | 60.5% | 0.0038 | Table S4 | Fig.7(b) |

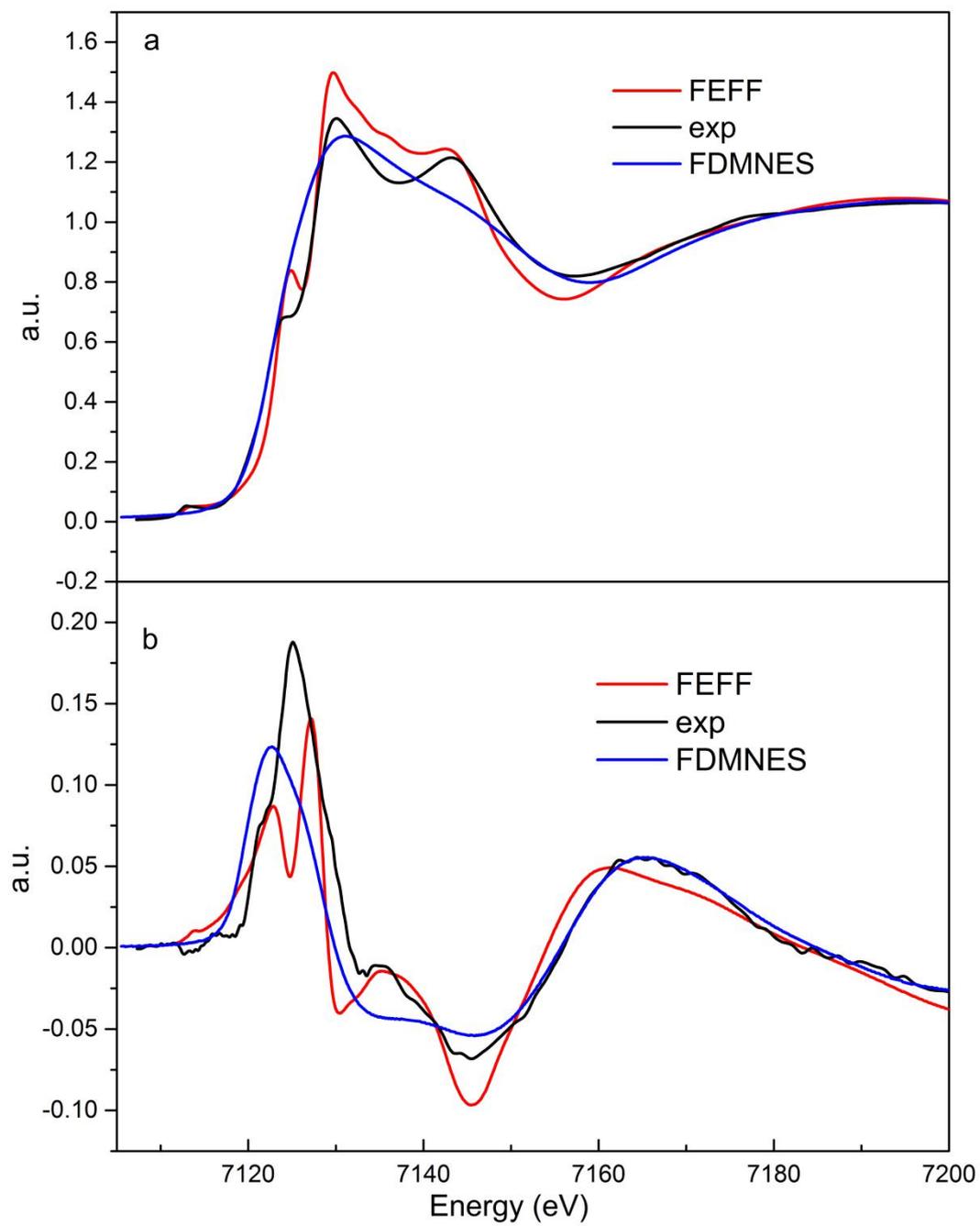

**Figure 7** (a) The ground state XANES calculation of FEFF/NOMAD and FDMNES/NOMAD, (b) the difference XANES calculation of FEFF/NOMAD(FEFF) and FDMNES/NOMAD(FDMNES)

**Table 3** Summary of the difference fitting spectrum compared with HS analogue result

|  | First shell bond length change(Å) | Intermediate fraction |
|---|---|---|
| HS analogue | +0.17 | 35% |
| Difference EXAFS(real and imaginary part) | +0.17 | 37% |
| Difference XANES | +0.11 | 39% |

## 4. Conclusion and Outlook

We developed a two-step scheme for data analysis of the TR-XAS. First we fit the difference EXAFS in R-space and get the bond length change of first shell coordination, then we fit difference XANES to get three dimensional structure. The scheme is flexible in XANES fit where both the spectrum calculation package and the global optimization algorithm for variables are changeable. This scheme was applied in the analysis of a photo-induced spin crossover iron complex. The fit results are in agreement with structure of the HS state analogue and the reported results. Next, some other XAFS calculation packages, such as 'xspectra' based on projector augmented wave function (Gougoussis *et al.*, 2009) and OCEAN based on Bethe-Salpeter equation (Gilmore *et al.*, 2015), will be included in our scheme.

## Acknowledgement

We thank Qingyuan Meng and Guowei Huang in Prof. Lizu Wu's group at Technical Institute of Physics and Chemistry for the synthesis of the Fe(II)(2-CH$_3$-phen)$_3$ and the advice on the Fe(II)(phen)$_3$ order. We are grateful to Xiaoyi Zhang at the Advanced Photon Source of the Argonne National Laboratory for the TR-XAS data collection. This work is supported by the National Natural Science Foundation of China (U1332205) and the Knowledge Innovation program of the Chinese Academy of Sciences (KJCX2-W-N42).

**Supporting information**

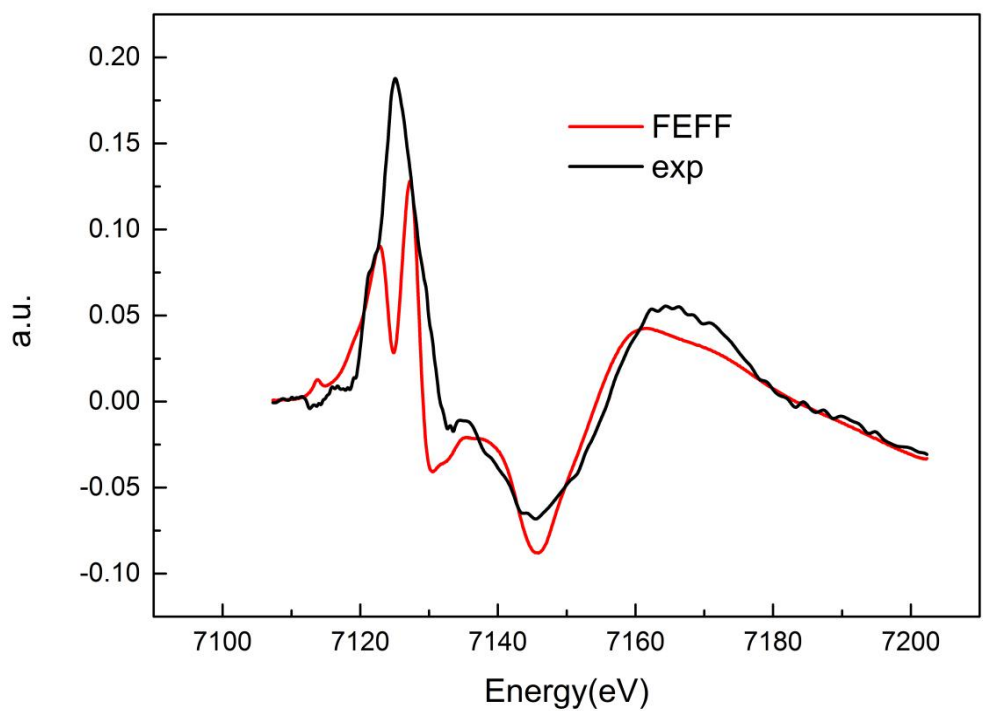

**Figure S1. Experimental difference XANES (exp) and fit result (FEFF) of HS state (FEFF**

**calculation with DIRECT_L optimization algorithm).**

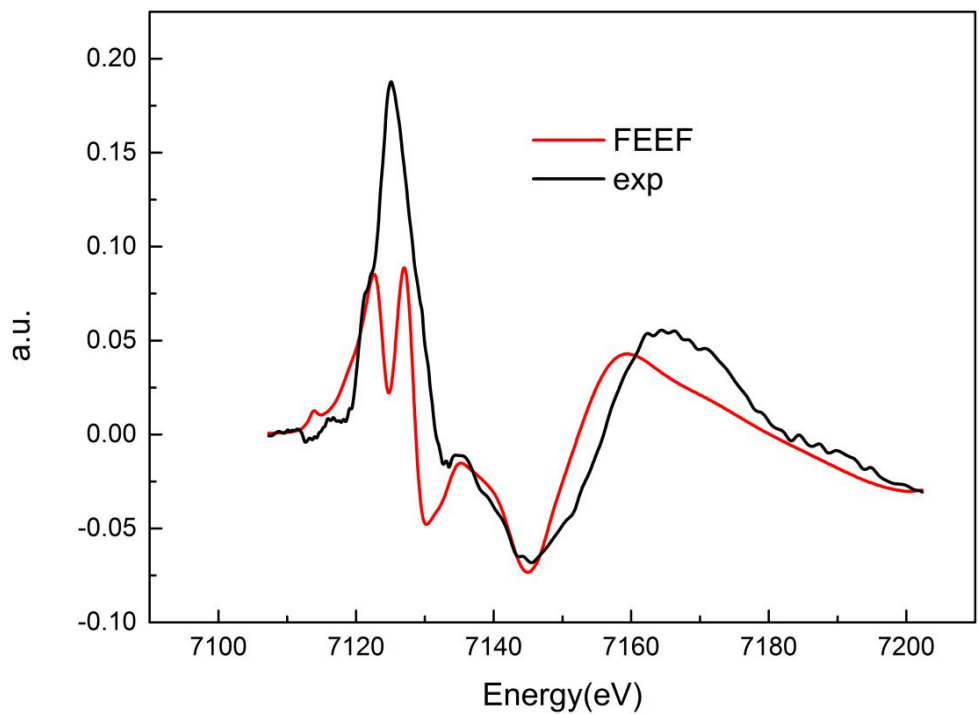

**Figure S2. Experimental difference XANES (exp) and fit result (FEFF) of HS state (FEFF calculation with ISRES optimization algorithm).**

**Table S1. Fit result of HS state (FEFF with NOMAD)**

The coordinates of atoms in Fe(II)(phen)$_3$ complex for HS state with best fit in unit Å

| Fe | 0.00000000 | 0.00000000 | 0.00000000 |
|----|-----------|-----------|-----------|
| N | 0.36600000 | 0.04300000 | 2.00400000 |
| N | -1.22200000 | -1.67000000 | -0.18700000 |
| N | 1.27000000 | -1.44000000 | -0.88400000 |
| N | 1.13200000 | 1.63700000 | -0.54900000 |
| N | -1.46800000 | 1.30200000 | 0.65100000 |
| N | -0.51600000 | 0.28000000 | -2.04200000 |
| C | -1.30500000 | 1.70400000 | 1.94700000 |
| C | -0.32000000 | 1.00500000 | 2.68900000 |
| C | -0.54400000 | -2.84900000 | -0.35900000 |
| C | 1.26800000 | -0.66200000 | 2.68000000 |
| H | 1.74900000 | -1.34200000 | 2.22400000 |
| C | 0.82600000 | -2.71400000 | -0.75100000 |
| C | 0.16200000 | 1.27900000 | -2.66800000 |
| C | 1.05400000 | 2.03000000 | -1.85400000 |
| C | -1.37200000 | -0.44600000 | -2.77200000 |
| H | -1.85800000 | -1.15000000 | -2.35600000 |
| C | -2.51500000 | -1.75500000 | 0.17700000 |
| H | -3.01600000 | -0.95800000 | 0.29200000 |
| C | -2.42000000 | 1.89800000 | -0.05500000 |
| H | -2.57200000 | 1.61700000 | -0.95000000 |
| C | 1.95100000 | 2.34000000 | 0.24900000 |
| H | 2.03400000 | 2.08000000 | 1.15800000 |
| C | 2.54900000 | -1.28600000 | -1.26400000 |
| H | 2.89000000 | -0.40500000 | -1.36400000 |
| C | -0.09700000 | 1.30900000 | 4.03800000 |
| C | 2.69100000 | 3.43900000 | -0.20500000 |
| H | 3.24100000 | 3.92400000 | 0.39800000 |
| C | 0.02000000 | 1.59500000 | -4.02400000 |
| C | 1.77800000 | 3.10400000 | -2.38900000 |
| C | -1.10300000 | -4.11100000 | -0.17200000 |
| C | -2.04800000 | 2.73400000 | 2.54500000 |
| C | 1.54100000 | -0.44100000 | 4.04100000 |
| H | 2.18600000 | -0.97200000 | 4.49300000 |
| C | -3.21500000 | 2.93100000 | 0.47200000 |
| H | -3.88400000 | 3.33700000 | -0.06700000 |
| C | 3.39400000 | -2.36400000 | -1.51300000 |
| H | 4.29000000 | -2.21000000 | -1.79000000 |
| C | 1.61000000 | -3.84800000 | -0.96900000 |
| C | -3.13600000 | -2.98500000 | 0.38900000 |
| H | -4.04700000 | -3.00900000 | 0.65600000 |
| C | -1.56700000 | -0.19100000 | -4.13700000 |
| H | -2.18100000 | -0.72100000 | -4.63000000 |
| C | 2.94100000 | -3.62700000 | -1.36400000 |
| H | 3.52100000 | -4.36300000 | -1.52300000 |
| C | 0.86800000 | 0.55100000 | 4.71300000 |
| H | 1.05500000 | 0.72200000 | 5.63000000 |
| C | -3.02700000 | 3.35100000 | 1.75500000 |
| H | -3.55600000 | 4.05600000 | 2.11000000 |
| C | -0.85300000 | 2.37300000 | 4.62800000 |
| H | -0.70200000 | 2.60500000 | 5.53800000 |
| C | -2.45400000 | -4.15500000 | 0.22100000 |
| H | -2.88500000 | -4.98900000 | 0.36600000 |
| C | 1.61100000 | 3.39700000 | -3.78400000 |
| H | 2.10300000 | 4.11200000 | -4.17300000 |
| C | -1.77500000 | 3.05000000 | 3.91300000 |
| H | -2.25600000 | 3.75500000 | 4.33000000 |
| C | 0.77000000 | 2.67800000 | -4.54900000 |
| H | 0.67700000 | 2.90100000 | -5.46700000 |
| C | -0.88500000 | 0.80600000 | -4.76200000 |
| H | -1.01900000 | 0.97100000 | -5.68900000 |
| C | 2.61700000 | 3.80800000 | -1.51100000 |
| H | 3.13200000 | 4.54100000 | -1.82800000 |
| C | 1.01600000 | -5.12100000 | -0.76600000 |
| H | 1.54400000 | -5.89900000 | -0.90300000 |
| C | -0.27300000 | -5.26400000 | -0.38500000 |
| H | -0.63400000 | -6.13500000 | -0.25700000 |

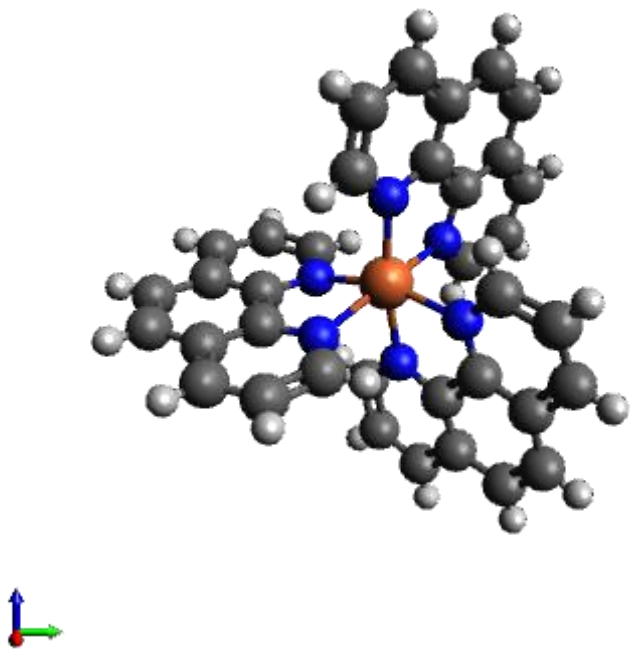

**Figure S3. Molecule of structure in Table S1 (FEFF/NOMAD).**

**Table S2. Fit result of HS state (FEFF calculation with DIRECT_L optimization algorithm)**

The coordinates of atoms in Fe(II)(phen)$_3$ complex for HS state with best fit in unit Å

| Fe | 0.00000000 | 0.00000000 | 0.00000000 |
|---|---|---|---|
| N | 0.45700000 | 0.08400000 | 1.96800000 |
| N | -1.12000000 | -1.72000000 | -0.09500000 |
| N | 1.37200000 | -1.49000000 | -0.79200000 |
| N | 1.20400000 | 1.54500000 | -0.48200000 |
| N | -1.37700000 | 1.34300000 | 0.61500000 |
| N | -0.44400000 | 0.18800000 | -1.97500000 |
| C | -1.21400000 | 1.74500000 | 1.91100000 |
| C | -0.22900000 | 1.04600000 | 2.65300000 |
| C | -0.44200000 | -2.89900000 | -0.26700000 |
| C | 1.35900000 | -0.62100000 | 2.64400000 |
| H | 1.84000000 | -1.30100000 | 2.18800000 |
| C | 0.92800000 | -2.76400000 | -0.65900000 |
| C | 0.23400000 | 1.18700000 | -2.60100000 |
| C | 1.12600000 | 1.93800000 | -1.78700000 |
| C | -1.30000000 | -0.53800000 | -2.70500000 |
| H | -1.78600000 | -1.24200000 | -2.28900000 |
| C | -2.41300000 | -1.80500000 | 0.26900000 |
| H | -2.91400000 | -1.00800000 | 0.38400000 |
| C | -2.32900000 | 1.93900000 | -0.09100000 |
| H | -2.48100000 | 1.65800000 | -0.98600000 |
| C | 2.02300000 | 2.24800000 | 0.31600000 |
| H | 2.10600000 | 1.98800000 | 1.22500000 |
| C | 2.65100000 | -1.33600000 | -1.17200000 |
| H | 2.99200000 | -0.45500000 | -1.27200000 |
| C | -0.00600000 | 1.35000000 | 4.00200000 |
| C | 2.76300000 | 3.34700000 | -0.13800000 |
| H | 3.31300000 | 3.83200000 | 0.46500000 |
| C | 0.09200000 | 1.50300000 | -3.95700000 |
| C | 1.85000000 | 3.01200000 | -2.32200000 |
| C | -1.00100000 | -4.16100000 | -0.08000000 |
| C | -1.95700000 | 2.77500000 | 2.50900000 |
| C | 1.63200000 | -0.40000000 | 4.00500000 |
| H | 2.27700000 | -0.93100000 | 4.45700000 |
| C | -3.12400000 | 2.97200000 | 0.43600000 |
| H | -3.79300000 | 3.37800000 | -0.10300000 |
| C | 3.49600000 | -2.41400000 | -1.42100000 |
| H | 4.39200000 | -2.26000000 | -1.69800000 |
| C | 1.71200000 | -3.89800000 | -0.87700000 |
| C | -3.03400000 | -3.03500000 | 0.48100000 |
| H | -3.94500000 | -3.05900000 | 0.74800000 |
| C | -1.49500000 | -0.28300000 | -4.07000000 |
| H | -2.10900000 | -0.81300000 | -4.56300000 |
| C | 3.04300000 | -3.67700000 | -1.27200000 |
| H | 3.62300000 | -4.41300000 | -1.43100000 |
| C | 0.95900000 | 0.59200000 | 4.67700000 |
| H | 1.14600000 | 0.76300000 | 5.59400000 |
| C | -2.93600000 | 3.39200000 | 1.71900000 |
| H | -3.46500000 | 4.09700000 | 2.07400000 |
| C | -0.76200000 | 2.41400000 | 4.59200000 |
| H | -0.61100000 | 2.64600000 | 5.50200000 |
| C | -2.35200000 | -4.20500000 | 0.31300000 |
| H | -2.78300000 | -5.03900000 | 0.45800000 |
| C | 1.68300000 | 3.30500000 | -3.71700000 |
| H | 2.17500000 | 4.02000000 | -4.10600000 |
| C | -1.68400000 | 3.09100000 | 3.87700000 |
| H | -2.16500000 | 3.79600000 | 4.29400000 |
| C | 0.84200000 | 2.58600000 | -4.48200000 |
| H | 0.74900000 | 2.80900000 | -5.40000000 |
| C | -0.81300000 | 0.71400000 | -4.69500000 |
| H | -0.94700000 | 0.87900000 | -5.62200000 |
| C | 2.68900000 | 3.71600000 | -1.44400000 |
| H | 3.20400000 | 4.44900000 | -1.76100000 |
| C | 1.11800000 | -5.17100000 | -0.67400000 |
| H | 1.64600000 | -5.94900000 | -0.81100000 |
| C | -0.17100000 | -5.31400000 | -0.29300000 |
| H | -0.53200000 | -6.18500000 | -0.16500000 |

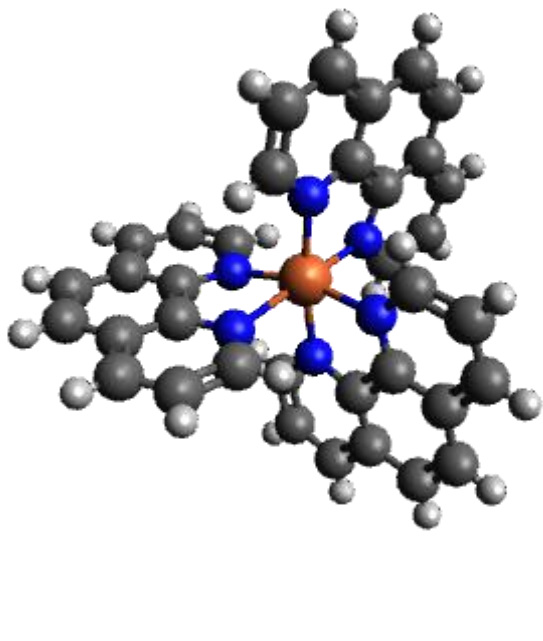

**Figure S4. Molecule of structure in Table S2.**

**Table S3. Fit result of HS state (FEFF calculation with ISRES optimization algorithm)**

The coordinates of atoms in Fe(II)(phen)$_3$ complex for HS state with best fit in unit Å

| Fe | 0.00000000 | 0.00000000 | 0.00000000 |
|---|---|---|---|
| N | 0.34600000 | 0.11600000 | 2.10700000 |
| N | -0.90400000 | -1.55700000 | 0.59900000 |
| N | 1.58800000 | -1.32700000 | -0.09800000 |
| N | 1.35200000 | 1.45000000 | -0.67800000 |
| N | -1.48800000 | 1.37500000 | 0.75400000 |
| N | -0.29600000 | 0.09300000 | -2.17100000 |
| C | -1.32500000 | 1.77700000 | 2.05000000 |
| C | -0.34000000 | 1.07800000 | 2.79200000 |
| C | -0.22600000 | -2.73600000 | 0.42700000 |
| C | 1.24800000 | -0.58900000 | 2.78300000 |
| H | 1.72900000 | -1.26900000 | 2.32700000 |
| C | 1.14400000 | -2.60100000 | 0.03500000 |
| C | 0.38200000 | 1.09200000 | -2.79700000 |
| C | 1.27400000 | 1.84300000 | -1.98300000 |
| C | -1.15200000 | -0.63300000 | -2.90100000 |
| H | -1.63800000 | -1.33700000 | -2.48500000 |
| C | -2.19700000 | -1.64200000 | 0.96300000 |
| H | -2.69800000 | -0.84500000 | 1.07800000 |
| C | -2.44000000 | 1.97100000 | 0.04800000 |
| H | -2.59200000 | 1.69000000 | -0.84700000 |
| C | 2.17100000 | 2.15300000 | 0.12000000 |
| H | 2.25400000 | 1.89300000 | 1.02900000 |
| C | 2.86700000 | -1.17300000 | -0.47800000 |
| H | 3.20800000 | -0.29200000 | -0.57800000 |
| C | -0.11700000 | 1.38200000 | 4.14100000 |
| C | 2.91100000 | 3.25200000 | -0.33400000 |
| H | 3.46100000 | 3.73700000 | 0.26900000 |
| C | 0.24000000 | 1.40800000 | -4.15300000 |
| C | 1.99800000 | 2.91700000 | -2.51800000 |
| C | -0.78500000 | -3.99800000 | 0.61400000 |
| C | -2.06800000 | 2.80700000 | 2.64800000 |
| C | 1.52100000 | -0.36800000 | 4.14400000 |
| H | 2.16600000 | -0.89900000 | 4.59600000 |
| C | -3.23500000 | 3.00400000 | 0.57500000 |
| H | -3.90400000 | 3.41000000 | 0.03600000 |
| C | 3.71200000 | -2.25100000 | -0.72700000 |
| H | 4.60800000 | -2.09700000 | -1.00400000 |
| C | 1.92800000 | -3.73500000 | -0.18300000 |
| C | -2.81800000 | -2.87200000 | 1.17500000 |
| H | -3.72900000 | -2.89600000 | 1.44200000 |
| C | -1.34700000 | -0.37800000 | -4.26600000 |
| H | -1.96100000 | -0.90800000 | -4.75900000 |
| C | 3.25900000 | -3.51400000 | -0.57800000 |
| H | 3.83900000 | -4.25000000 | -0.73700000 |
| C | 0.84800000 | 0.62400000 | 4.81600000 |
| H | 1.03500000 | 0.79500000 | 5.73300000 |
| C | -3.04700000 | 3.42400000 | 1.85800000 |
| H | -3.57600000 | 4.12900000 | 2.21300000 |
| C | -0.87300000 | 2.44600000 | 4.73100000 |
| H | -0.72200000 | 2.67800000 | 5.64100000 |
| C | -2.13600000 | -4.04200000 | 1.00700000 |
| H | -2.56700000 | -4.87600000 | 1.15200000 |
| C | 1.83100000 | 3.21000000 | -3.91300000 |
| H | 2.32300000 | 3.92500000 | -4.30200000 |
| C | -1.79500000 | 3.12300000 | 4.01600000 |
| H | -2.27600000 | 3.82800000 | 4.43300000 |
| C | 0.99000000 | 2.49100000 | -4.67800000 |
| H | 0.89700000 | 2.71400000 | -5.59600000 |
| C | -0.66500000 | 0.61900000 | -4.89100000 |
| H | -0.79900000 | 0.78400000 | -5.81800000 |
| C | 2.83700000 | 3.62100000 | -1.64000000 |
| H | 3.35200000 | 4.35400000 | -1.95700000 |
| C | 1.33400000 | -5.00800000 | 0.02000000 |
| H | 1.86200000 | -5.78600000 | -0.11700000 |
| C | 0.04500000 | -5.15100000 | 0.40100000 |
| H | -0.31600000 | -6.02200000 | 0.52900000 |

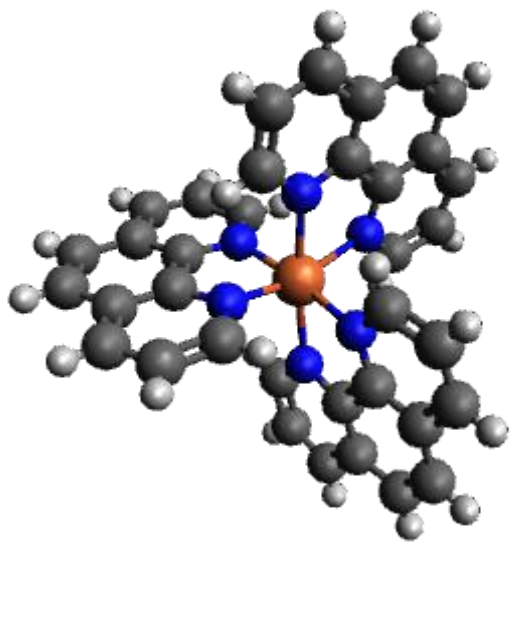

**Figure S5. Molecule of structure in Table S3.**

**Table S4. Fit result of HS state (FDMNES calculation with NOMAD optimization algorithm)**

The coordinates of atoms in Fe(II)(phen)₃ complex for HS state with best fit in unit Å

| Fe | 0.00000000 | 0.00000000 | 0.00000000 |
|----|------------|------------|------------|
| N | 0.41700000 | 0.14500000 | 2.07900000 |
| N | -1.11500000 | -1.77900000 | 0.40500000 |
| N | 1.37700000 | -1.54900000 | -0.29200000 |
| N | 1.19000000 | 1.51300000 | -0.42500000 |
| N | -1.41700000 | 1.40400000 | 0.72600000 |
| N | -0.45800000 | 0.15600000 | -1.91800000 |
| C | -1.25400000 | 1.80600000 | 2.02200000 |
| C | -0.26900000 | 1.10700000 | 2.76400000 |
| C | -0.43700000 | -2.95800000 | 0.23300000 |
| C | 1.31900000 | -0.56000000 | 2.75500000 |
| H | 1.80000000 | -1.24000000 | 2.29900000 |
| C | 0.93300000 | -2.82300000 | -0.15900000 |
| C | 0.22000000 | 1.15500000 | -2.54400000 |
| C | 1.11200000 | 1.90600000 | -1.73000000 |
| C | -1.31400000 | -0.57000000 | -2.64800000 |
| H | -1.80000000 | -1.27400000 | -2.23200000 |
| C | -2.40800000 | -1.86400000 | 0.76900000 |
| H | -2.90900000 | -1.06700000 | 0.88400000 |
| C | -2.36900000 | 2.00000000 | 0.02000000 |
| H | -2.52100000 | 1.71900000 | -0.87500000 |
| C | 2.00900000 | 2.21600000 | 0.37300000 |
| H | 2.09200000 | 1.95600000 | 1.28200000 |
| C | 2.65600000 | -1.39500000 | -0.67200000 |
| H | 2.99700000 | -0.51400000 | -0.77200000 |
| C | -0.04600000 | 1.41100000 | 4.11300000 |
| C | 2.74900000 | 3.31500000 | -0.08100000 |
| H | 3.29900000 | 3.80000000 | 0.52200000 |
| C | 0.07800000 | 1.47100000 | -3.90000000 |

```
C            1.83600000     2.98000000    -2.26500000

C           -0.99600000    -4.22000000     0.42000000

C           -1.99700000     2.83600000     2.62000000

C            1.59200000    -0.33900000     4.11600000

H            2.23700000    -0.87000000     4.56800000

C           -3.16400000     3.03300000     0.54700000

H           -3.83300000     3.43900000     0.00800000

C            3.50100000    -2.47300000    -0.92100000

H            4.39700000    -2.31900000    -1.19800000

C            1.71700000    -3.95700000    -0.37700000

C           -3.02900000    -3.09400000     0.98100000

H           -3.94000000    -3.11800000     1.24800000

C           -1.50900000    -0.31500000    -4.01300000

H           -2.12300000    -0.84500000    -4.50600000

C            3.04800000    -3.73600000    -0.77200000

H            3.62800000    -4.47200000    -0.93100000

C            0.91900000     0.65300000     4.78800000

H            1.10600000     0.82400000     5.70500000

C           -2.97600000     3.45300000     1.83000000

H           -3.50500000     4.15800000     2.18500000

C           -0.80200000     2.47500000     4.70300000
```

| | | | |
|---|---|---|---|
| H | -0.65100000 | 2.70700000 | 5.61300000 |
| C | -2.34700000 | -4.26400000 | 0.81300000 |
| H | -2.77800000 | -5.09800000 | 0.95800000 |
| C | 1.66900000 | 3.27300000 | -3.66000000 |
| H | 2.16100000 | 3.98800000 | -4.04900000 |
| C | -1.72400000 | 3.15200000 | 3.98800000 |
| H | -2.20500000 | 3.85700000 | 4.40500000 |
| C | 0.82800000 | 2.55400000 | -4.42500000 |
| H | 0.73500000 | 2.77700000 | -5.34300000 |
| C | -0.82700000 | 0.68200000 | -4.63800000 |
| H | -0.96100000 | 0.84700000 | -5.56500000 |
| C | 2.67500000 | 3.68400000 | -1.38700000 |
| H | 3.19000000 | 4.41700000 | -1.70400000 |
| C | 1.12300000 | -5.23000000 | -0.17400000 |
| H | 1.65100000 | -6.00800000 | -0.31100000 |
| C | -0.16600000 | -5.37300000 | 0.20700000 |
| H | -0.52700000 | -6.24400000 | 0.33500000 |

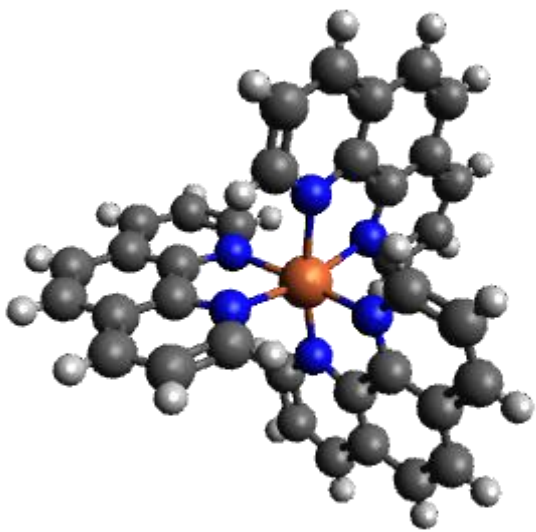

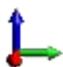

**Figure S6. Molecule of structure in Table S4**